\documentclass[tc]{copernicus}
\pdfoutput=1
\usepackage{xcolor}
\title{Calving on tidewater glaciers amplified by submarine frontal melting}
\author[1,2]{M. O'Leary}
\author[1]{P. Christoffersen}
\affil[1]{Scott Polar Research Institute, University of Cambridge, Cambridge, UK}
\affil[2]{Department of Atmospheric, Oceanic and Space Sciences, University of Michigan, Ann Arbor, MI, USA}
\runningtitle{Calving amplified by frontal melting}
\runningauthor{M. O'Leary and P. Christoffersen}
\correspondence{Martin O'Leary (olearym@umich.edu)}
\begin{document}

\maketitle
\begin{abstract}
While it has been shown repeatedly that ocean conditions exhibit an important control on the behaviour of grounded tidewater glaciers, modelling studies have focused largely on the effects of basal and surface melting. Here, a finite-element model of stresses near the front of a tidewater glacier is used to investigate the effects of frontal melting on calving, independently of the calving criterion used. Applications of the stress model to idealized scenarios reveal that undercutting of the ice front due to frontal melting can drive calving at up to ten times the mean melt rate. Factors which cause increased frontal melt-driven calving include a strong thermal gradient in the ice, and a concentration of frontal melt at the base of the glacier. These properties are typical of both Arctic and Antarctic tidewater glaciers. The finding that frontal melt near the base is a strong driver of calving leads to the conclusion that water temperatures near the bed of the glacier are critically important to the glacier front, and thus the flow of the glacier. These conclusions are robust against changes in the basal boundary condition and the choice of calving criterion, as well as variations in the glacier size or level of crevassing.
\end{abstract}

\introduction

The calving of icebergs makes up a large component of the mass balance of many polar ice sheets and glaciers \citep{Hagen2003Net,Rignot2006Changes}. Recent observations of increased calving and coincident flow acceleration at a number of tidewater glaciers \citep{Joughin2004Large,Luckman2006Rapid,Stearns2007Rapid,Howat2007Rapid} have shown that calving and glacier flow are strongly linked. However, the modelling of calving processes is still problematic, leading to a great deal of uncertainty in predictions of the future behaviour of tidewater glaciers, and their consequent contributions to sea-level rise.

Reviewing the mechanisms of calving at tidewater glaciers, \citet{Benn2007Review} distinguish between first-order calving mechanisms associated with longitudinal stretching of the glacier, and second-order mechanisms associated with buoyant forcing at the front. A calving law based on simplified physics \citep{Benn2007Calving} has shown some success in replicating these first-order processes \citep{Mottram2009Testing, Nick2010Physically, Cook2012Testing}. However, no model exists in current practice to account for the second-order processes, despite the possibility that these may account for the majority of calving in some circumstances.

It was first noted by \citet{Weertman1957Deformation} that there is a necessary imbalance at any ice front between the glaciostatic pressure outwards and the hydrostatic pressure inwards. In floating ice this imbalance manifests itself as a pure bending moment acting on the ice front, while in grounded ice it partially expresses itself as a net outward force. \citet{Reeh1968Calving} demonstrated the consequences of this effect on an analytic model of an ice shelf with a Newtonian rheology. He showed that a maximum in both tensile stress and surface elevation develops about one ice thickness from the front, and that this stress leads to calving, with the calving rate determined by the glacier's thickness and viscosity. He also noted the possible effects of variations in the shape of the ice front, although these were not incorporated into his model.

This same stress maximum has been described by a number of authors. \citet{Fastook1982Finite} showed the same effect in a finite-element model of a Newtonian floating ice shelf, while \citet{Hanson2000Glacier} demonstrated a similar stress maximum in a grounded model with a non-linear rheology. \citet{Scambos2009Ice} showed similar results for the flexure of an ice shelf, using realistic temperature profiles as inputs to a Glen-type rheology.

While the idea that frontal melting could be a driver of calving has been mentioned by a number of authors \citep{Hanson2000Glacier,Vieli2002Retreat}, there have been few quantitative studies of the phenomenon. A number of authors \citep{Kirkbride1997Calving,Benn2001Growth,Rohl2006ThermoErosional} have identified the melting and erosion of waterline notches in lake-terminating glaciers as controls on their calving rates. Similarly, estimates of frontal melting at LeConte Glacier, Alaska \citep{Motyka2003Submarine} and several Greenlandic glaciers \citep{Rignot2010Rapid, Sutherland2012Estimating} show that melting may be an important term in the frontal mass balance of these glaciers. \citet{Jenkins2011ConvectionDriven} has provided some explanation of these results, using a model of plume-driven melting. Given the evidence for the importance of frontal melt as a mass balance term, and the likelihood of a connection with calving processes, it has become a subject of interest how tidewater glaciers react to melt-driven changes in their front geometry.

In this study a simple model of two-dimensional laminar ice flow is described, and its results are used to investigate the response of a tidewater glacier to submarine frontal melting. Section (\ref{sec:flowmodel}) outlines the model equations, boundary conditions and implementation. The theoretical framework in which the results are presented is described in Sections (\ref{sec:retreat}) and (\ref{sec:multiplier}), which introduce the key ideas of stress retreat and wet calving multiplier, and uses model results to show that the latter of these is well-founded. Finally, Section (\ref{sec:ae}) shows how these can be used to predict the changes in calving behaviour that should be expected under particular conditions of frontal melt.

\section{Ice flow model}
\label{sec:flowmodel}
\subsection{Flow of ice}

While ice exhibits visco-elastic properties on hourly to daily time-scales, the motion of glaciers over time-scales longer than about a day is best described as that of viscous fluid flow \citep{Paterson2000Physics}. As the Reynolds numbers involved are extremely low, of the order of 10\textsuperscript{-13}, and the ice density is constant throughout most of the glacier body, the flow can be approximated using the incompressible Stokes equations:

\begin{eqnarray}
\nabla p &=& \mu \nabla^2 u + f \\
\nabla \cdot u &=& 0
\end{eqnarray}

Here $p$ is the scalar pressure, $u$ is the flow velocity, $f$ is the body force (in this case gravity), and $\mu$ is the dynamic viscosity of the fluid. As ice exhibits non-Newtonian flow properties, $\mu$ is itself a function of the velocity field, and must be specified through a flow law.

The most common flow law in use for glaciological purposes is that derived by \citet{Glen1952Experiments}. This is most simply expressed as a relationship between the stress and strain tensors. The deviatoric stress tensor is denoted here by $\tau_{ij}$, and the second invariant of this tensor calculated as:

\begin{equation}
  2\tau^2 = \sum_{i,j} \tau_{ij}^2
\end{equation}

The strain rates $\dot \epsilon_{ij} = \frac{1}{2}\left(\frac{\partial u_i}{\partial x_j} + \frac{\partial u_j}{\partial x_i}\right)$ can then be expressed as a function of the deviatoric stresses:

\begin{equation}
  \dot \epsilon_{ij} = A \tau^{n - 1} \tau_{ij}
  \label{eqn:glen}
\end{equation}

The quantities $A$ and $n$ are considered parameters of the flow law. In this work $n$ is fixed at its conventional value of 3, and $A$ is allowed to vary as a function of temperature $T$:

\begin{equation}
  A = A_0 \exp \left( -Q/RT \right)
\end{equation}

The universal gas constant $R$ is fixed, while the activation energy $Q$ and the multiplicative factor $A_0$ are chosen to match measurements. For the purposes of this study, standard values are used for $Q$ and $A_0$, as given by \citet{Paterson2000Physics}.

To translate between the stress-strain relationship of Equation (\ref{eqn:glen}) and the viscosity-based Stokes formulation, Glen's law is inverted, giving the effective viscosity in terms of the strain rates and the flow parameter $A$:

\begin{equation}
  \mu = A^{-1/n} \left( \sum_{i,j} \dot \epsilon_{ij}^2 \right)^\frac{1-n}{2n}
  \label{eqn:effvis}
\end{equation}

\subsection{Boundary conditions}

On the upper surface of the glacier, as well as that portion of the front which is above the waterline, a stress-free boundary condition is applied. Below the waterline, hydrostatic pressure is applied. At the rear of the domain, far from the front, the flow velocity is set to a constant of 1 km a\textsuperscript{-1}. Sensitivity tests (not shown here) indicate that the results are largely insensitive to this value.

The basal boundary condition for glacier flow is an active topic of research, and several relationships have been proposed. The most common of these are those based on the work of \citet{Weertman1957Sliding}, who gives a power law relationship between basal shear stress $\tau_b$, basal velocity $u_b$, and the effective pressure at the base $N = p_i - p_w$:

\begin{equation}
  \tau_b = k u_b^p N^{-q}
\end{equation}

This relationship, while convenient numerically, and difficult to disprove empirically, has been shown to have difficulties, notably the lack of an upper bound for shear stress. A calculation by \citet{Iken1981Effect} showed that for a bounded basal slope, such a bound must exist. This result was reproduced in a more general setting by \citet{Schoof2005Effect}, who suggested a relationship which was refined by \citet{Gagliardini2007Finiteelement}:

\begin{equation}
  \tau_b = C N \left(1 + \frac{\lambda^* A C^n N^n}{m^* u} \right)^{-1/n}
  \label{eqn:schoof}
\end{equation}

Here $\lambda^*$ and $m^*$ are the dominant wavelength and slope of bed features, and $C$ is a constant subject to the inequality $C \le m^*$. Following \citet{Pimentel2010Hydrologically} the relationship $C = 0.84 m^*$ is used, based on the result for a sinusoidal bed. We assume that water pressure is hydrostatic, based on a free connection to the ocean.

\subsection{Numerics}

For the purposes of this study, a two-dimensional solution to the incompressible Stokes equations is sought, using Glen's flow law as a constitutive relation. The glacier is treated as of uniform width, and sufficiently wide that lateral boundary effects are unimportant.

Such a solution is produced using the free open-source finite element solver FreeFem++ \citep{Hecht2005Freefem}, using a standard triangular P2 element for the velocity field and a P1 element for the pressure, both implemented on an unstructured grid. A finite element implementation allows the model to easily handle a variety of geometries, as well as allowing the model's resolution to be focused on the areas of interest. In order to handle the implicit definition of $\mu$ through Equation (\ref{eqn:effvis}), the system is solved iteratively, beginning with the Newtonian solution.

The basal boundary condition (Equation (\ref{eqn:schoof})) is also non-linear, which presents some difficulties in the numerical implementation. A Robin-type boundary condition is used, expressing $\tau_b$ as a linear multiple of $u$ at the base, and recalculating the constant of proportionality with each step of the non-linear iteration procedure. This iterative process is combined with that for the effective viscosity, in order to reduce the total number of iterations required.

This procedure is found to be much more numerically stable than the alternatives, such as Dirichlet or Neumann boundary conditions, and it does not substantially increase the number of iterations required for convergence over the case of a fixed basal velocity. 

\section{Stress retreat}
\label{sec:retreat}
\subsection{Assumptions}

The aim of this study is to quantify the effects of submarine frontal melting on calving rate. To this end, some assumptions must be made about the calving
criterion. It is assumed that calving behaviour is determined solely by the viscous properties of ice, neglecting any effects due to elastic deformation. This greatly simplifies the calculation of stresses, and is in line with most glaciological practice. 

It is also assumed that the variable of greatest interest is the (Cauchy) stress, rather than the strain rate (or equivalently, deviatoric stress). This is supported by \cite{Vaughan1993Relating}, who notes that strain rates at crevasse sites vary by almost three orders of magnitude while stresses are almost constant. For simplicity, only the first (most tensile) principal stress is considered, as it is assumed to be the controlling factor on fracture. External sources of stress are ignored, based on the analysis of \cite{Bassis2008Investigation}.

It is known \citep{Weertman1973Can} that water-filled crevasses are likely to penetrate the full thickness of the glacier, under most realistic stress conditions. Therefore, if we assume that crevasses which reach below the waterline are likely to contain water, we can disregard the stresses which occur at depths much below the waterline, for the purposes of calculating crevasse depths. 

Similarly, if we assume that the dominant control on the crevasse's growth is the stress field around the crack tip, the stress field near to the surface of the glacier is irrelevant, once the crack has grown beyond a certain size. Thus, assuming that compressive stress increases with depth, the most important factor in determining whether the crevasse penetrates the whole glacier is the stress field around the waterline.

It should be noted that the Benn-Nye calving criterion \citep{Benn2007Calving}, as widely used in models of tidewater glaciers, fits all of the above assumptions, as do any obvious modifications of it.

Finally, it is assumed that internal deformation of the glacier takes place on timescales much longer than those associated with calving or frontal melting. This follows from the calculation of \citet{vanderVeen2002Calving}, who showed that a typical tidewater glacier front deforms at a rate which is between one and two orders of magnitude too slow to be responsible for calving.

It therefore suffices to look only at the cumulative amount of frontal melt, rather than the rate at which it occurs. Another advantage of this assumption is that it can be internally verified by using the stress field to calculate instantaneous velocities. Note that no assumption is made about the absolute velocity of the glacier, merely about the rate at which it is deforming internally. In fact, for a stable calving front it is necessary that the glacier sliding velocity is of a similar order to the calving rate, and thus must take place on the same timescale.

\subsection{Definitions}

In this study, the dimensionless dry calving length $\varepsilon$ is defined as the aspect ratio of the incipient iceberg, when no frontal melt is occurring. Thus, for a glacier of thickness $H$, the distance from the ice front to the point of calving is $\varepsilon H$. While this measure is undoubtedly variable, and may even include a stochastic component, it is fixed for a given calving event. It is also useful to define the undercut length $d$ as the depth-averaged cumulative frontal melt, and the dimensionless waterline height $h$ as the ratio of the glacier depth below sea level to the full glacier thickness.

\begin{figure}[t]
\vspace*{2mm}
\begin{center}
\includegraphics[width=8.3cm]{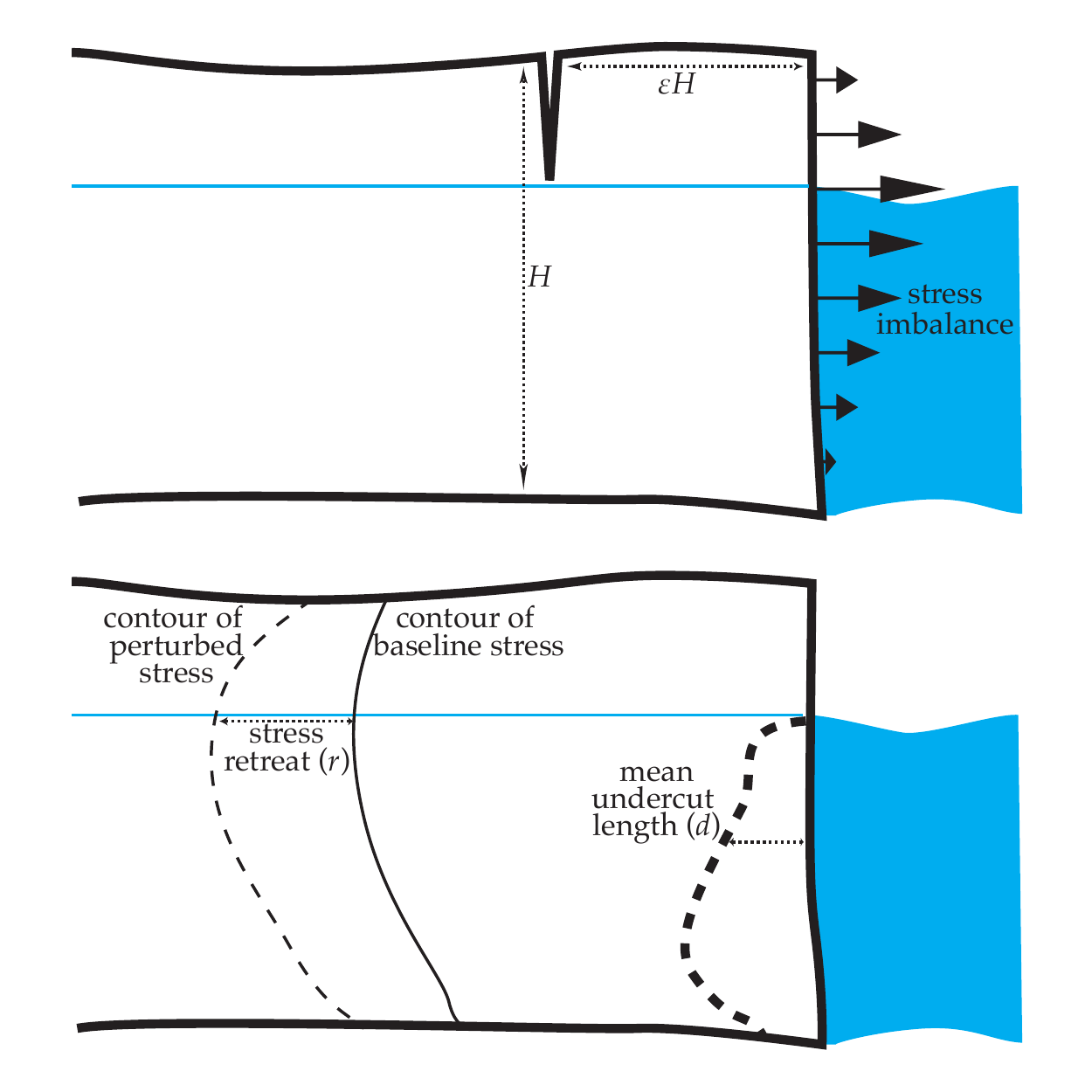}
\caption{Above: Schematic of the geometry used in this study. Below: Schematic of change in stress due to an undercut. The stress retreat $r$ is the distance between a stress contour in the baseline state, and the equivalent contour in the perturbed state, measured along the waterline.}
\label{fig:stressretreat}
\end{center}	
\end{figure}

The central concept of this analysis is the `stress retreat', which is a measure of the spatial effect of a stress perturbation, defined as follows. A reference frame is used whereby $x$ is a horizontal variable, increasing inland from zero at the ice front. Given a reference stress state $\sigma_{ref}(x)$ and a perturbed state $\sigma_{pert}(x)$, the stress retreat $r$ is the minimum distance inland such that $\sigma_{ref}(x) = \sigma_{pert}(x + r)$. In other words, the effect of the perturbation is to move the stress field inland by a distance $r$. This distance is, of course, a function of position $x$, and may vary considerably. In all cases considered here, $r$ is positive and finite.

Intuitively, the stress retreat can be thought of as the distance that the stress field is `pushed back' by a perturbation, such as undercutting or a change in the frontal boundary condition. Ice in this situation will behave as though it were this distance further forward in an unperturbed glacier.

Finally, the wet calving multiplier $\omega$ is defined as $r/d$, where $r$ is the stress retreat due to an undercut length $d$. From first principles, there is no reason to suppose that $\omega$ is independent of $d$, but this shall be shown empirically to be the case in Section (\ref{sec:multiplier}).

The ratio $\omega$ can be interpreted as follows: assuming, in the absence of frontal melting, that the conditions for calving exist at a point $x$, then after a quantity of frontal melting $d$, those same conditions exist at the point $x + r = x + \omega d$. Thus, if calving is occurring at an interval $\Delta t$, leading to a dry calving rate of $x/\Delta t$, the calving rate once frontal melting is incorporated will be equal to $x/\Delta t + \omega d/\Delta t$, an increase of $\omega$ times the mean melt rate. 

It may therefore be useful to treat $\omega$ as a measure of the sensitivity of the calving rate to variations in frontal melt rate, with a value of one corresponding to a simple additive model. Such a way of thinking about the effects of submarine melt is the dominant one in those works which include melting in the frontal mass balance calculation \citep{Motyka2003Submarine,Amundson2010Unifying}. While such an approach is observationally correct --- there is no easy way to distinguish frontal melt-driven calving from any other kind --- it will be found more productive here to make this distinction, and to separate `dry' melt-free calving from `wet' frontal melt-driven calving.

This interpretation must be used with care, however. An increase in the mean size of a calving event is likely to have repercussions on the dynamics of the glacier. As such, the effects described here should be considered merely one component of a system of interacting processes and feedbacks which ultimately determine the behaviour of the glacier.

\section{Wet calving multiplier}
\label{sec:multiplier}

In order to calculate the effects of undercutting on near-frontal stress, the flow model is used to compute stress fields in a variety of configurations. As a baseline unperturbed model run, meant to simulate a typical medium-sized tidewater glacier, the model is run in a flat rectangular slab configuration. The ice thickness is 300 m, and the basal parameters are $\lambda^* =$ 20 m and $m^* =$ 0.13 --- see below for the sensitivities to these parameters. The water level varies between model runs, as the results are quite sensitive to this variable.

\begin{figure}[t]
\vspace*{2mm}
\begin{center}
\includegraphics[width=8.3cm]{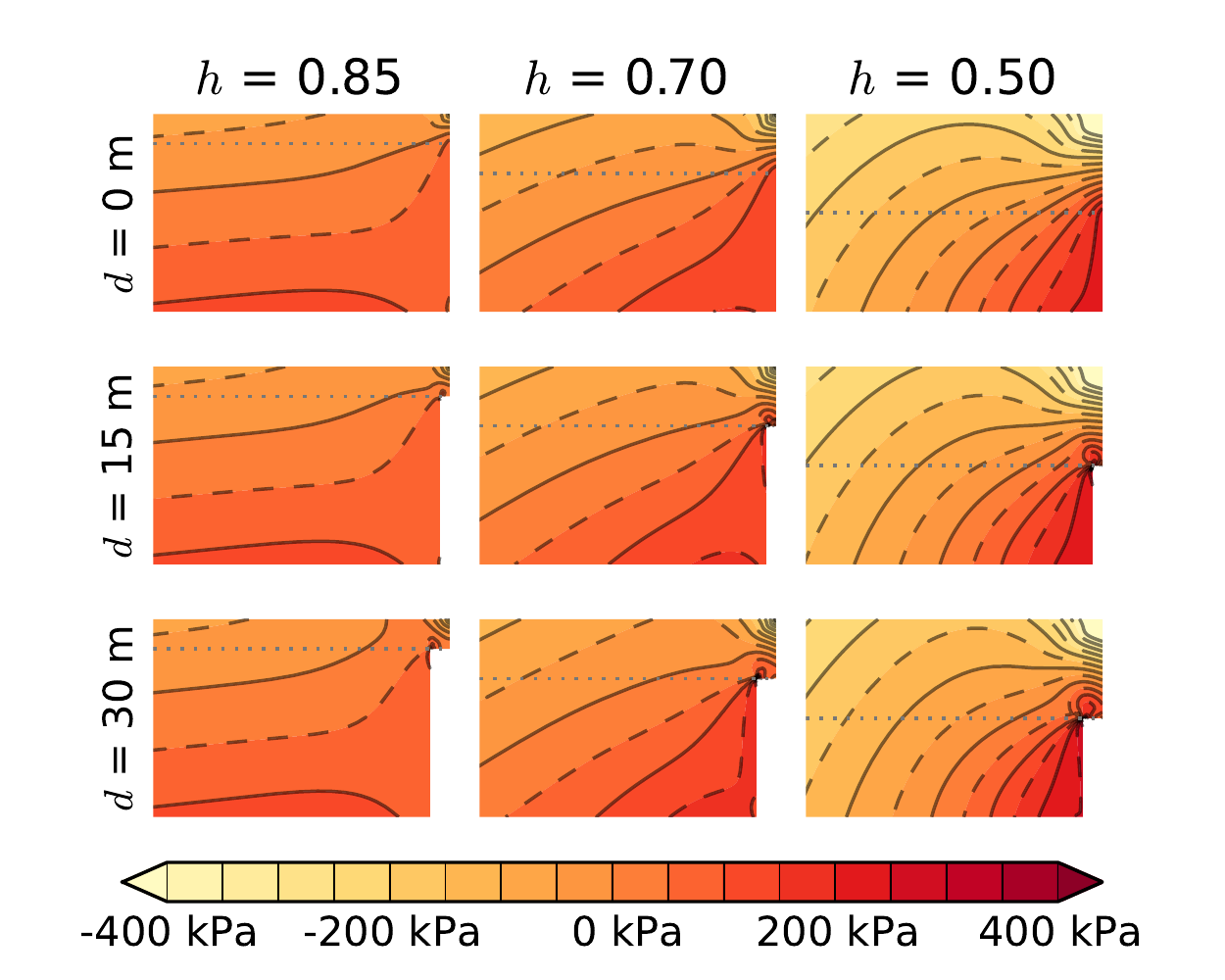}
\caption{Comparison of first principal Cauchy stresses, relative to hydrostatic pressure, for a variety of water depths and undercut lengths, assuming uniform undercutting. Grey dashed line indicates water level.}
\label{fig:stressgrid}
\end{center}	
\end{figure}

By altering the shape of the domain, the effect of undercutting by frontal melt can be simulated. Figure (\ref{fig:stressgrid}) shows the deviatoric stress fields generated by a selection of geometries and water levels, assuming uniform frontal melt below sea level, and an isothermal glacier. Qualitatively, it seems clear that undercutting results in an increase in tension due to the bending moment exerted by the overhang, as well as the reduction in basal traction near the glacier foot. These effects increase with the undercut length $d$ and show qualitative variation with the water level $h$.

\begin{figure}[t]
\vspace*{2mm}
\begin{center}
\includegraphics[width=8.3cm]{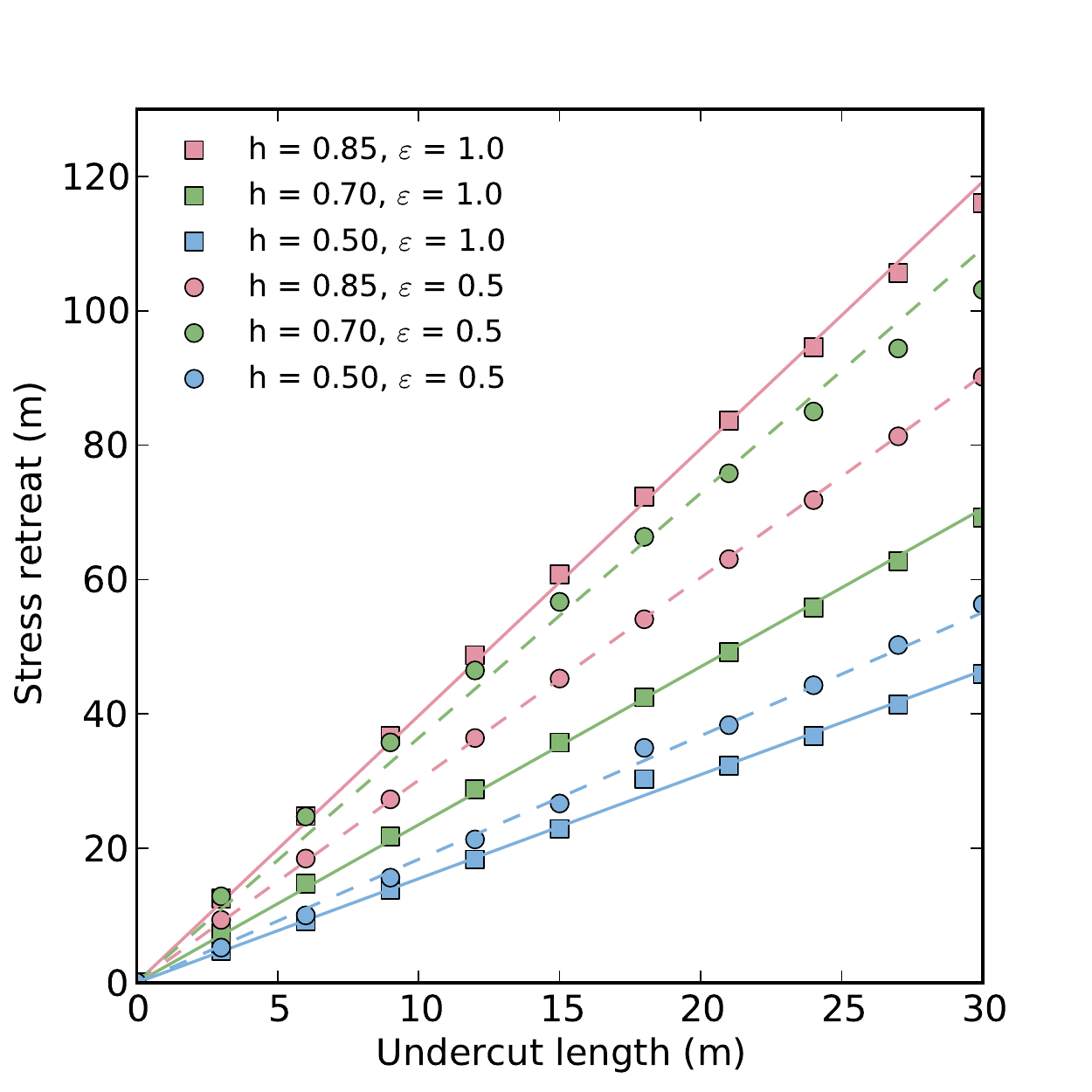}
\caption{Stress retreat as a function of undercut length for a variety of scenarios involving different water depths $h$ and dry calving lengths $\varepsilon$. In all cases the relationship is very close to linear. The slope of the linear fit is $\omega$, the wet calving multiplier.}
\label{fig:compression}
\end{center}	
\end{figure}

Figure (\ref{fig:compression}) shows the stress retreat as a function of undercut length for a variety of scenarios. In each case, a uniform undercut is introduced and the stress retreat measured, relative to a particular dry calving length $\varepsilon$. For each scenario, a linear fit through the origin is possible, with $R^2 > 0.99$. The slope of this fit is equal to the wet calving multiplier $\omega$, which is henceforth assumed to be independent of the undercut length. Although this assumption must certainly break down at large undercut lengths, it appears to hold for undercut lengths which are up to 20\% of an ice thickness, far larger than the expected depth of real-world undercuts.

\section{Sensitivities}
\label{sec:ae}
Given that $\omega$ is well-defined for a given scenario, the question remains as to what factors influence its value. The most obvious of these is the dry calving length $\varepsilon$. As this is used as the initial point from which the stress retreat is measured, it should come as no surprise that the magnitude of the stress retreat (and hence $\omega$) is dependent on its value.

For a grounded or partially grounded glacier, values of $\varepsilon$ greater than one are usually held to be unlikely, given that the resulting berg would be unable to capsize, and would thus have no obvious route of escape from the glacier. Here, the upper limit is drawn at a value of $\varepsilon = 1.5$, to allow for some leeway in the system. Similarly, values of $\varepsilon < 0.25$ are neglected, as such narrow calving events are likely to be much more affected by the detailed geometry of the front than by viscous stresses. However, it should be noted that they are not ruled out by this model, merely likely to be modelled incorrectly.

Another variable of interest is the water depth, or more loosely, the `degree of grounding'. As there is known to be a significant difference in calving behaviour between grounded and floating glaciers \citep{Walter2010Iceberg}, it seems reasonable to suggest that the water level may have a significant qualitative effect on calving, even if the transition is not as abrupt as that between grounded and floating ice. As such, a selection of water depths are investigated, ranging from $h = 0.5$, for a well-grounded glacier, to $h = 0.85$, a glacier almost at the point of flotation. Flotation occurs at $h = \rho_i/\rho_w \simeq 0.89$, at which point the calculations of the model diverge in a non-useful manner.

\begin{figure}[t]
\vspace*{2mm}
\begin{center}
\includegraphics[width=8.3cm]{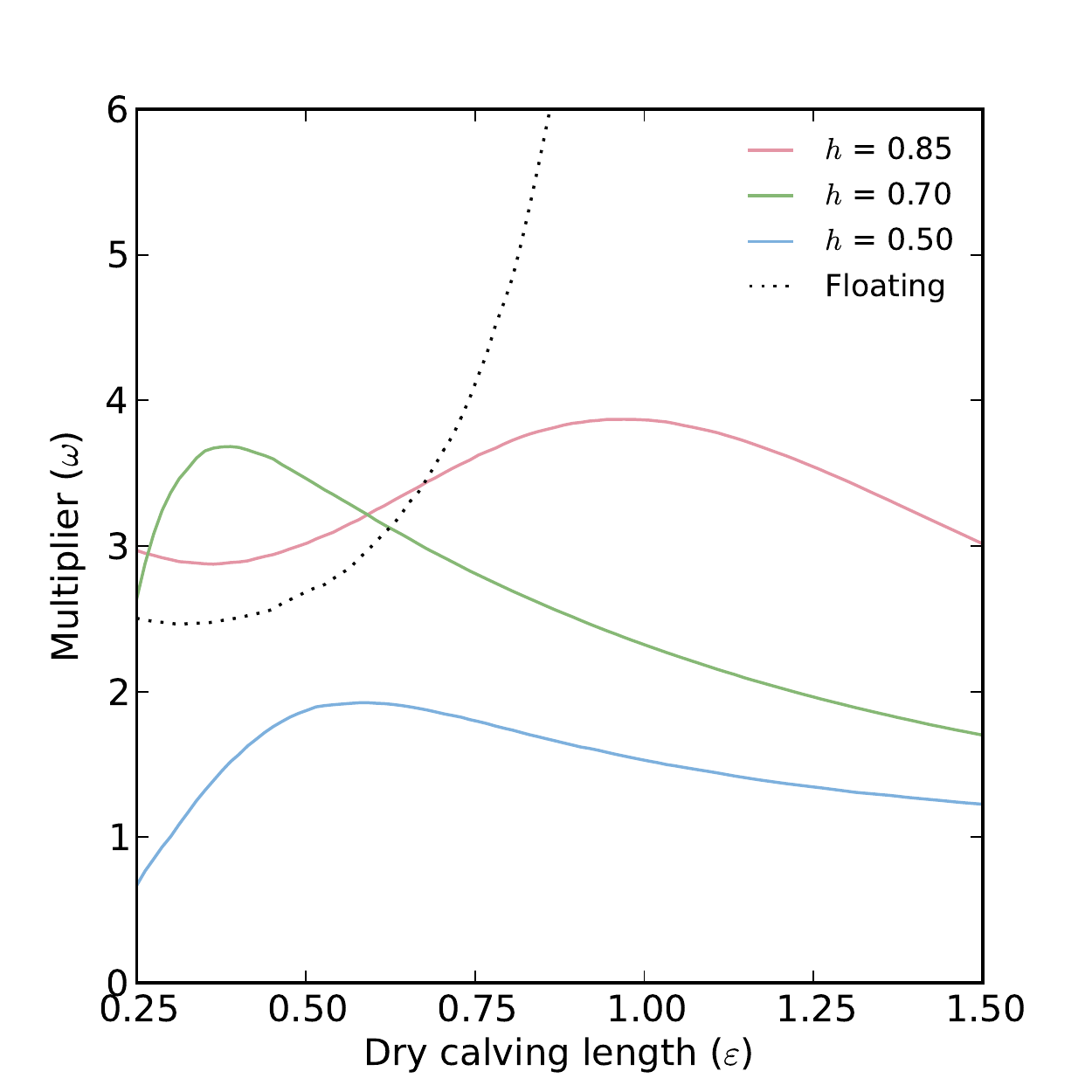}
\caption{Relationship of wet calving multiplier $\omega$ to calving length $\varepsilon$ for varying water depths. Also shown is the floating case, where $\omega$ rapidly diverges.}
\label{fig:ae_basic}
\end{center}	
\end{figure}

As can be seen in Figure (\ref{fig:ae_basic}), typical values of $\omega$ for the simplest scenarios are in the range 1--4, indicating that in this idealised situation, frontal melt drives calving at a rate up to four times the mean melt rate. Higher water levels generally lead to larger values of the multiplier $\omega$, as do shorter calving lengths. In the $h=0.5$ case, where the glacier is immersed in water to its midpoint, the multiplier effect is relatively weak, with frontal melt-driven calving occurring at around one and a half times the melt rate. The effect is much stronger in the more typical $h = 0.7$ case, with a multiplier in the range of two to three. 

The highest values of $\omega$ are usually found at $\varepsilon \simeq 1 - h$, indicating that the geometry of the above-water portion of the nascent iceberg is important. At calving lengths shorter than this, there is some drop-off in the value of $\omega$, although as previously stated, the model may not be fully capturing the complexities of the stress field so close to the front.

Glaciers close to flotation also show an increase in $\omega$ for calving lengths of around an ice thickness, foreshadowing the divergence in $\omega$ when the glacier comes afloat. However care should be taken in interpreting this result, as the stress distribution becomes very uniform here, and in reality stress variations from other sources will probably play more of a role.

Significantly, in all but a few cases, $\omega$ is greater than one. This means that undercutting could be driving calving at rates greater than the frontal melt rate itself, providing an amplification of the oceanographic forcing on the glacier. This effect will be most strongly felt on those glaciers which are themselves well immersed in water. This may go some way towards explaining the empirical relationships which have been identified between calving and water depth \citep{Sikonia1982Finite,vanderVeen1996Tidewater}.

\subsection{Thermal regime and viscosity}

Given that many ice sheets are far from isothermal \citep{Paterson2000Physics}, it might be assumed that the thermal regime of their outlet glaciers is similarly heterogeneous. Although mechanisms have been suggested by which the hydrological system of a glacier could result in near-isothermal conditions \citep{Phillips2010CryoHydrologic}, this has not been widely observed in the field. As such, it is useful to consider the effects of a thermal gradient on the frontal melt-calving relationship.

\begin{figure}[t]
\vspace*{2mm}
\begin{center}
\includegraphics[width=8.3cm]{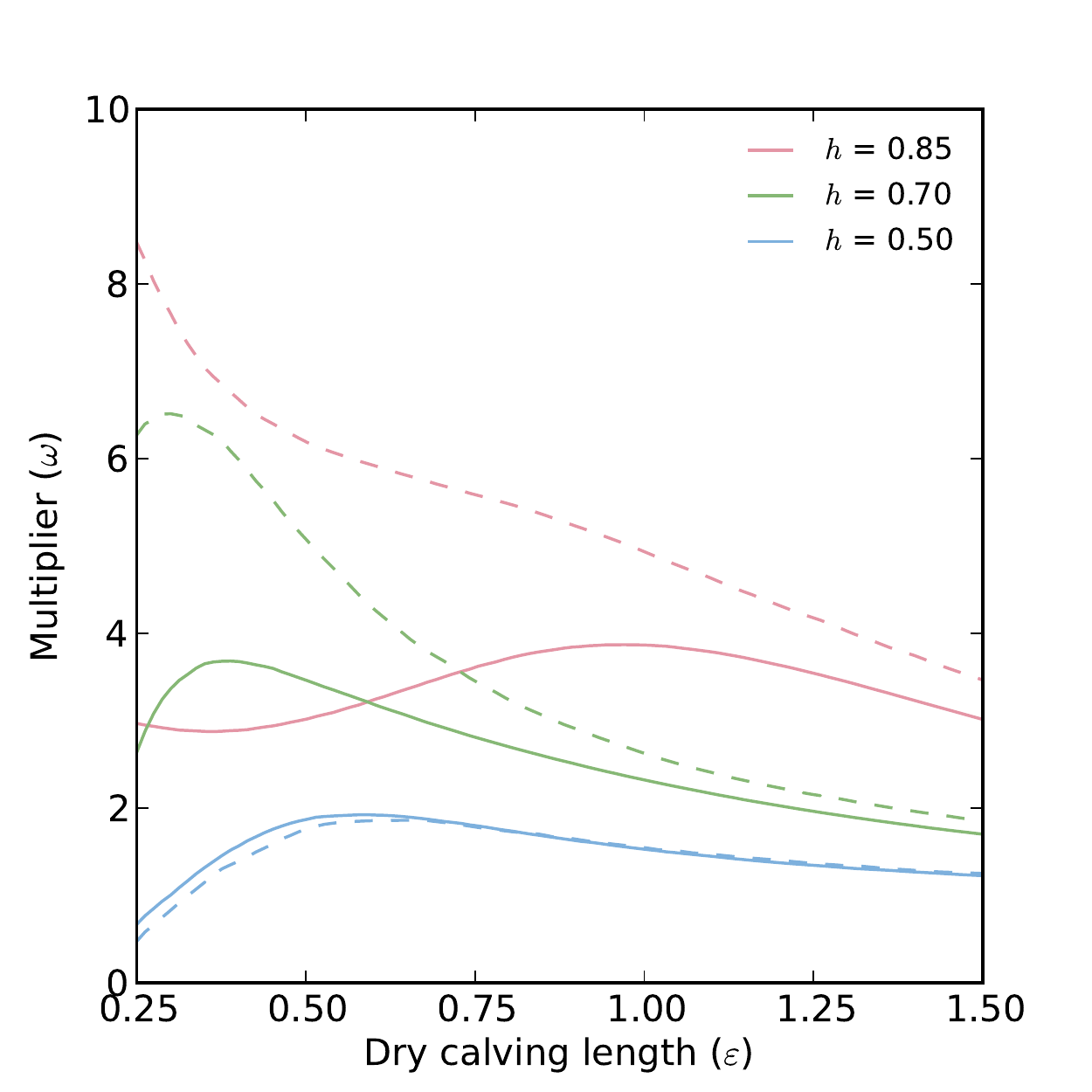}
\caption{The effect of a temperature gradient on the $\omega$-$\varepsilon$ relationship. Solid lines show the baseline cases, while dashed lines show the relationship when a 10\,\unit{\degree C} temperature gradient is applied. For lightly grounded glaciers, there is a positive effect, while heavily grounded glaciers see almost no difference.}
\label{fig:ae_grads}
\end{center}	
\end{figure}

For simplicity, a linear temperature trend within the ice is assumed, ranging from -10\,\unit{\degree C} at the surface to 0\,\unit{\degree C} at the bed. The effect of pressure on the melting point is neglected for simplicity. The effects are shown in Figure (\ref{fig:ae_grads}). In general, for glaciers close to flotation, the effect of the temperature gradient is to boost $\omega$, in some cases by around a factor of two. For more heavily grounded glaciers, such as the $h = 0.5$ case, the effect is very slightly negative.

The implication here is that glaciers in colder regions, with less developed hydrological systems and thus less homogeneous temperature profiles, are likely to be more severely impacted by a constant degree of undercutting, although these glaciers are also likely to be in colder marine environments, which will tend to decrease the quantity of submarine melt. In fact, higher atmospheric temperatures and the consequent surface melt may act to stabilize glacier fronts, by reducing the effects of undercutting on calving, if surface meltwater is sufficient to render the glacier temperate near the front.

It is likely that a two-part classification scheme is necessary here, distinguishing both between polar and temperate glaciers as is usual, but also between those terminating in warm and cold water. While there is certainly a strong correlation between the two groups, the relationship is imperfect, and there are certainly examples in Greenland of `cold' glaciers coming in contact with relatively warm water \citep{Rignot2010Rapid, Sutherland2012Estimating, ChristoffersenPartitioning}. It has also been shown \citep{Seale2011Ocean} that the behaviour of tidewater glaciers in Greenland respects strong geographical boundaries, which are more associated with oceanic than atmospheric or glacial temperatures.

\begin{figure}[tbp]
\vspace*{2mm}
\begin{center}
\includegraphics[width=8.3cm]{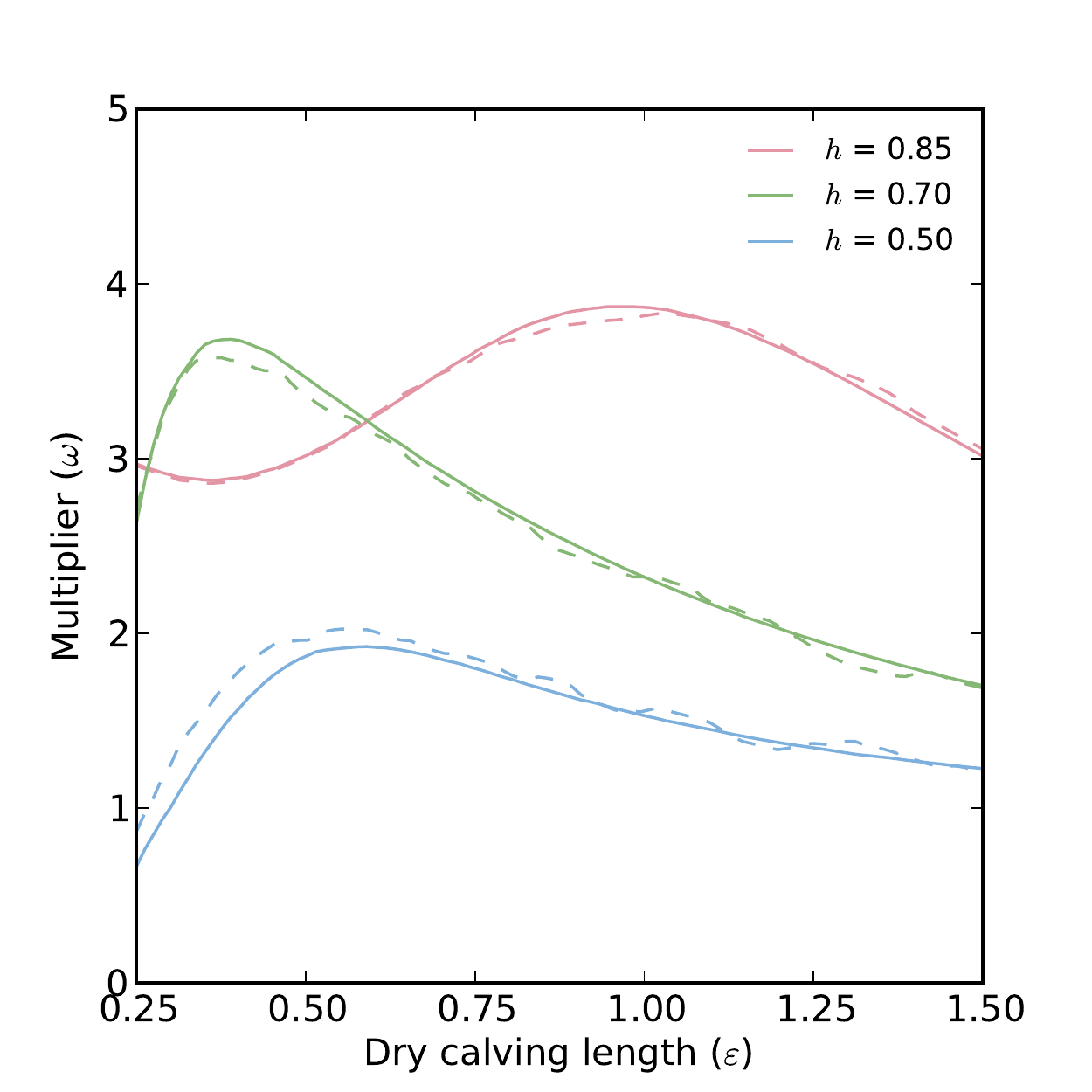}
\caption{The effect of surface damage on the $\omega$-$\varepsilon$ relationship. Dashed lines show the relationship when an enhancement factor is applied to the upper layers. No discernible pattern is visible, and the difference is likely to be a numerical artifact.}
\label{fig:ae_damaged}
\end{center}	
\end{figure}

Up until now, the glacier has been treated as a completely solid block of ice. In reality, the upper surface of a tidewater glacier is often heavily crevassed, resulting in reduced strength in the upper layers. Crevassing such as this is often represented in flow models through the use of an enhancement factor, which reduces the effective viscosity of the ice in such areas. Now, an enhancement factor is applied, reducing the effective viscosity above the waterline. At the surface the viscosity is reduced by a factor of ten, and this reduction is scaled linearly with depth until the viscosity reaches its normal value, 5m above the waterline. This height is chosen so as to minimise the effects of a `kink' in the effective viscosity at the waterline, which interferes with the accurate calculation of stresses.

The effects of this change can be seen in Figure (\ref{fig:ae_damaged}). While there are some small differences, it is quite likely that these are due to numerical inaccuracies, due in large part to the `kink' in the effective viscosity near the waterline. In general, it seems that the rheology of the above-water portion of the glacier has very little effect on the value of $\omega$. This is a very helpful result, as it provides some validation for the approach of using an idealized slab glacier model.

\subsection{Basal boundary condition}

\begin{figure}[t]
\vspace*{2mm}
\begin{center}
\includegraphics[width=8.3cm]{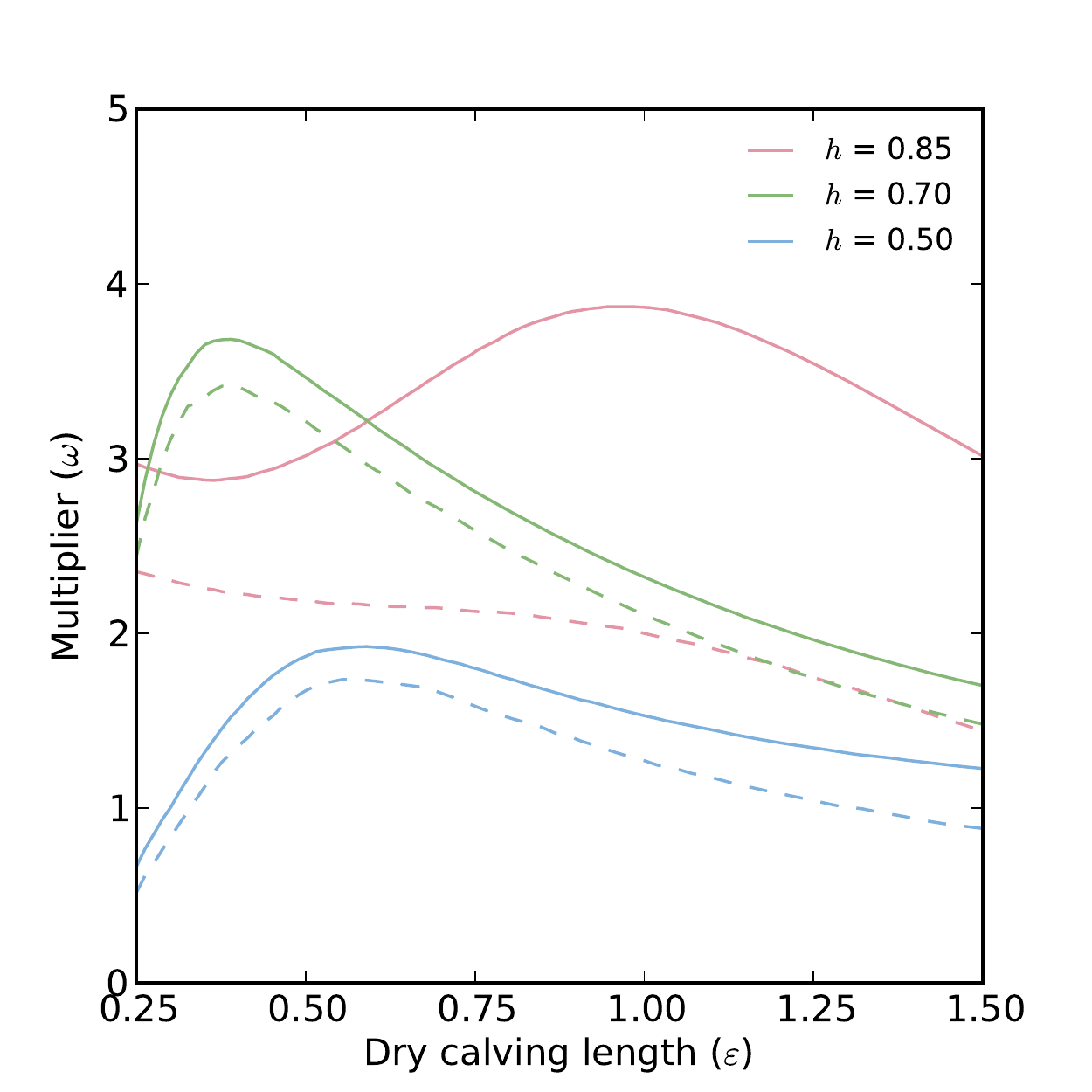}
\caption{The effect of a change in basal boundary condition on the $\omega$-$\varepsilon$ relationship. Dashed lines show the relationship with an extremely simple linear basal boundary condition. The most significant change is the reduction in $\omega$ for the $h = 0.85$ case.}
\label{fig:ae_basal}
\end{center}	
\end{figure}

The basal boundary condition is one of the most uncertain elements in any glacier model. Therefore, it seems prudent to check that the results given here are robust against changes in this area. As a simple test, the boundary condition given in Equation (\ref{eqn:schoof}) is replaced with a simple linear relationship between sliding velocity and basal traction, with the constant of proportionality (50 Pa a m\textsuperscript{-1}) chosen to best match the velocities of the more complex model.

The results are shown in Figure (\ref{fig:ae_basal}). The largest effect is the reduction in the value of $\omega$ for the nearly floating case, as the changes in effective pressure at the bed no longer can have any effect. In the other cases, there is a small decrease in the value of $\omega$, but the general pattern remains very similar. Therefore it can be concluded that the basal boundary condition has little effect on this particular calving mechanism.

Another possible source of variability is the glacier thickness itself. Although it can be argued through dimensional analysis that most of the physics should be unchanged by an increase or decrease in physical scale, there are some effects surrounding the basal boundary condition which do not scale in a simple way. However, sensitivity tests (not shown) indicate that the glacier thickness is unimportant, with the resulting graphs being visually indistinguishable. As such, the results given here can be thought of as independent of glacier thickness.

\subsection{Undercut shape}

Another consideration which has to this point been ignored is the vertical distribution of melt over the glacier face. As estimates of frontal melt rates have generally arisen from heat balance calculations, there is no empirical evidence to suggest a particular form for the frontal melt profile. However, given the modelling results of \citet{Jenkins2011ConvectionDriven}, it should be expected that the frontal melt rate would be vertically inhomogeneous.

An attempt can be made to quantify the effect of this inhomogeneity by using a variety of different idealized frontal melt profiles. As well as the uniform profile used thus far, the model can be run using undercuts in wedge shapes, as well as a parabolic curve which peaks in the center of the submarine ice face. Note that in all cases, the undercut length used in calculations is the mean melt rate on the ice face. This is equivalent to half the maximum frontal melt rate for the wedge shapes, or two thirds that in the parabolic case.

\begin{figure}[t]
\vspace*{2mm}
\begin{center}
\includegraphics[width=8.3cm]{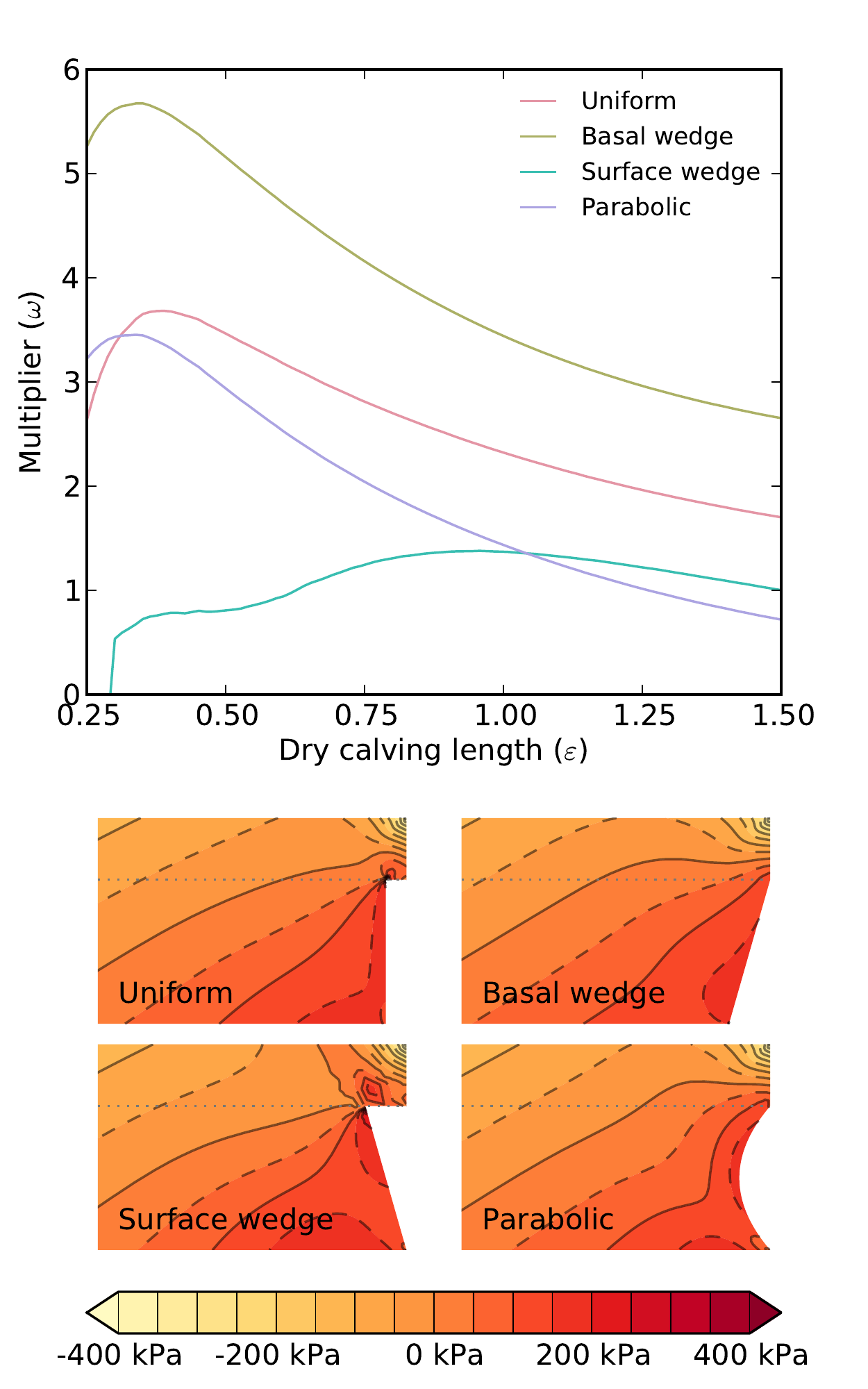}
\caption{Above: The $\omega$-$\varepsilon$ relationship for a variety of frontal melt profiles. Below: Cauchy stress fields (relative to hydrostatic pressure) associated with frontal melt profiles.}
\label{fig:ae_profiles}
\end{center}	
\end{figure}

Figure (\ref{fig:ae_profiles}) shows the results of these tests. The differences are large. The basal wedge results in values of $\omega$ which are consistently about 50\% greater than those in the uniform case. The surface wedge, by contrast, results in values which barely get above one, meaning that in this case, the simple additive approach is sufficient. Interestingly, the parabolic profile has a similar value of $\omega$ to the uniform case for small values of $\varepsilon$, but it drops off much more quickly for larger values.

Looking at the stress distributions, it can be suggested that the reason for the large value of $\omega$ in the basal wedge case is the movement of the `fulcrum' about which the glacier is bending to the base of the glacier. This area is visible in Figure (\ref{fig:ae_profiles}) as a region of high deviatoric stress. While the surface wedge case includes much higher stresses, they occur at the waterline and thus have only local effects.

\subsection{Surface slope}

A related issue to that of the undercut geometry is the geometry of the above-water portion of the ice, or equivalently the surface slope of the glacier. For simplicity of modelling, and due to the extremely complex and variable nature of real glacier geometries, we restrict our attention here to constant surface slopes.

\begin{figure}[t]
\vspace*{2mm}
\begin{center}
\includegraphics[width=8.3cm]{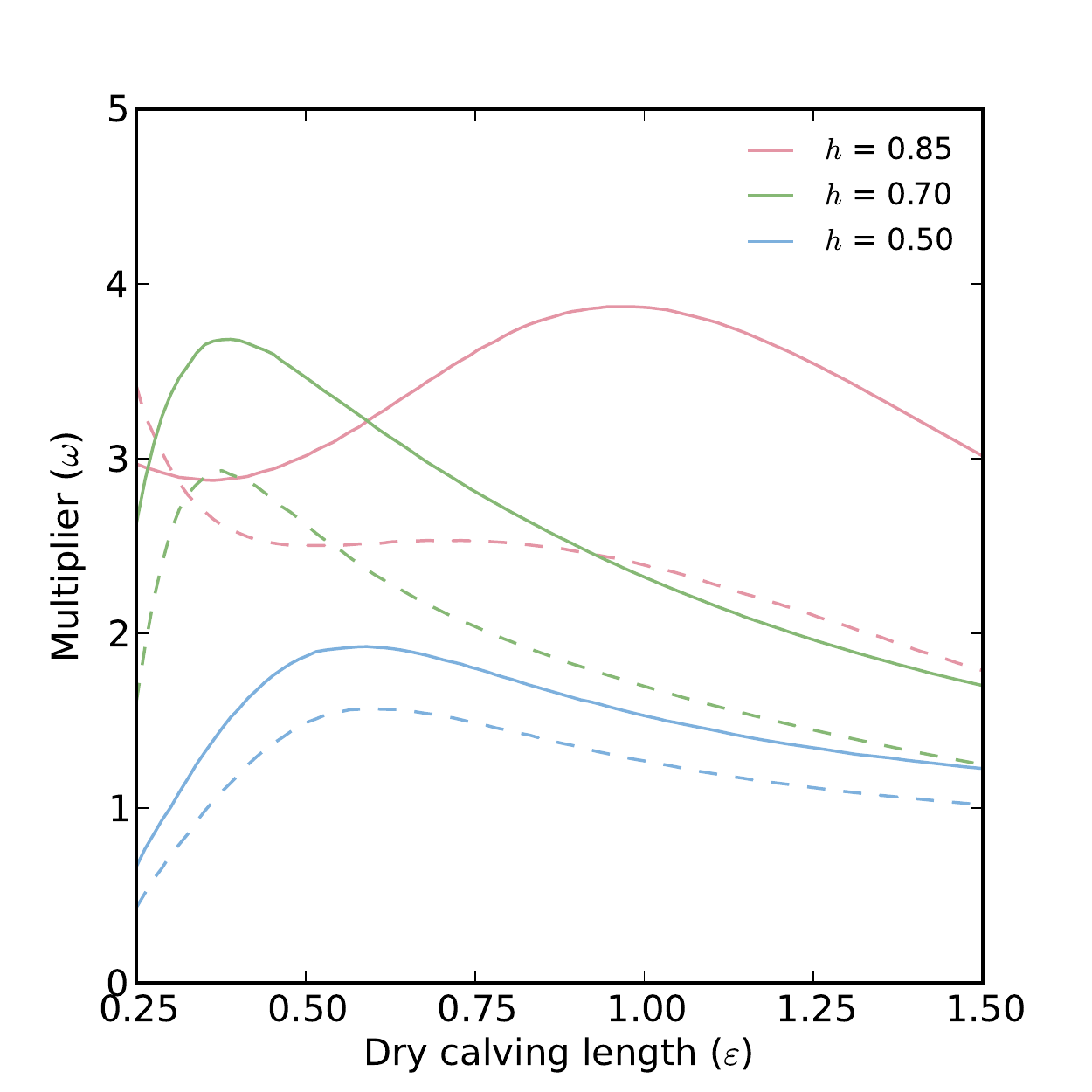}
\caption{The effect of a change in surface slope on the $\omega$-$\varepsilon$ relationship. Dashed lines show the relationship with a surface slope of 5\,\unit{\degree}. Note that in this case, $h$ refers to the waterline height at the ice front only, with effectively smaller values of $h$ further from the front}
\label{fig:ae_slope}
\end{center}	
\end{figure}

Figure (\ref{fig:ae_slope}) shows the effect of a 5\,\unit{\degree} surface slope on the calving multiplier. As might be expected, this has the effect of reducing the multiplier, relative to a glacier with a flat surface, as the sloped glacier is effectively `more grounded' inward of the front, a situation which we have shown results in a reduced multiplier. However, the effect is still significant, particularly at short calving lengths.

In the event of a reverse surface slope, as observed at Helheim and Kangerdlugssuaq by \citet{Joughin2008Icefront}, we should expect a similarly reversed effect on the multiplier. This is likely to lead to enhanced calving at longer lengths, as was observed by that study.

Owing to numerical simplifications in our model, we are unable to investigate the effects of a bed slope. While we believe these effects are likely to be quantitatively significant, we expect that they will not change the main qualitative conclusions of this study. Future work, applying this approach to real-world glacier conditions, will likely need to incorporate such effects.

\conclusions

The model described here is a novel ultra-high-resolution finite-element two-dimensional flowline model of the stresses near an ice front, incorporating state-of-the-art boundary conditions and full-Stokes ice physics. It provides a rapid and flexible means to investigate the stress fields near the calving fronts of tidewater glaciers. Using a simple set of assumptions about the form of a calving law, a theoretical framework has been developed which allows for conclusions to be drawn about the effect of a frontal perturbation on calving. It is shown that the effects of submarine frontal melting can be described within this framework, and that the impacts on calving may be much greater than previously assumed.

The model demonstrates that with uniform submarine melt, extra calving is generated at a rate between one and four times the melt rate. This effect is greater the closer a glacier is to flotation, and can be exacerbated by the existence of a thermal gradient within the glacier, or melting which is focused near the base of the ice front. Given the results of \citet{Jenkins2011ConvectionDriven}, which show that this form of frontal melting is quite likely on many glaciers, this result is of particular importance.

These results provide more evidence that the effects of water temperature near the base of tidewater glaciers are critical to the behaviour of the ice fronts. These findings are robust against both the basal boundary condition and the choice of calving law, and appear to be independent of the glacier size, as well as the level of damage present at the glacier surface.

\begin{acknowledgements}
  
This work was supported through a NERC Doctoral Training Grant to M. O'Leary. M. O'Leary was partially supported through NASA grant NNX08AN59G and NSF grant ARC1064535. We are grateful to N. Arnold, D. Benn and J. Bassis for their comments on this work, which have been very helpful in the development of the manuscript. We also acknowledge the assistance of the editor, O. Gagliardini, whose comments greatly improved this work.

\end{acknowledgements}

\bibliographystyle{copernicus}
\bibliography{mewo2}

\end{document}